\documentclass[twocolumn, switch]{article} 

\usepackage{aaai20}

\usepackage{amsmath, amsthm, amssymb, amsfonts}

\usepackage[numbers,square]{natbib}

\usepackage[utf8]{inputenc}	
\usepackage[T1]{fontenc}	
\usepackage{xcolor}		
\usepackage[colorlinks = true,
            linkcolor = purple,
            urlcolor  = blue,
            citecolor = cyan,
            anchorcolor = black]{hyperref}	
\usepackage{booktabs} 		
\usepackage{nicefrac}		
\usepackage{microtype}		
\usepackage{lineno}		
\usepackage{float}			
\usepackage{bbm}
\usepackage{amsmath}
\usepackage[noend, ruled]{algorithm2e}
\usepackage{times}
\usepackage{helvet}
\usepackage{courier}
\usepackage{physics}
\usepackage{graphicx}
\usepackage{lipsum}		

\usepackage{newfloat}
\DeclareFloatingEnvironment[name={Supplementary Figure}]{suppfigure}
\usepackage{sidecap}
\sidecaptionvpos{figure}{c}

\usepackage{titlesec}
\titlespacing\section{0pt}{12pt plus 3pt minus 3pt}{1pt plus 1pt minus 1pt}
\titlespacing\subsection{0pt}{10pt plus 3pt minus 3pt}{1pt plus 1pt minus 1pt}
\titlespacing\subsubsection{0pt}{8pt plus 3pt minus 3pt}{1pt plus 1pt minus 1pt}

\title{Reinforcement Learning with Quantum Variational Circuits}

\usepackage{xwatermark}
\newwatermark[firstpage,color=gray!90,angle=0,scale=0.28, xpos=0in,ypos=-5in]{*correspondence: \texttt{lockwo@rpi.edu}}

\usepackage{authblk}

\author[1\thanks{\tt{lockwo@rpi.edu}}]{Owen Lockwood}
\author[2]{Mei Si}
\affil[1]{Department of Computer Science, Rensselaer Polytechnic Institute}
\affil[2]{Department of Cognitive Science, Rensselaer Polytechnic Institute}

\begin{document}

\twocolumn[ 
  
\maketitle

\begin{abstract}
The development of quantum computational techniques has advanced greatly in recent years, parallel to the advancements in techniques for deep reinforcement learning. This work explores the potential for quantum computing to facilitate reinforcement learning problems. Quantum computing approaches offer important potential improvements in time and space complexity over traditional algorithms because of its ability to exploit the quantum phenomena of superposition and entanglement. Specifically, we investigate the use of quantum variational circuits, a form of quantum machine learning. We present our techniques for encoding classical data for a quantum variational circuit, we further explore pure and hybrid quantum algorithms for DQN and Double DQN. Our results indicate both hybrid and pure quantum variational circuit have the ability to solve reinforcement learning tasks with a smaller parameter space. These comparison are conducted with two OpenAI Gym environments: CartPole and Blackjack, The success of this work is indicative of a strong future relationship between quantum machine learning and deep reinforcement learning. 
\end{abstract}
\vspace{0.35cm}

] 



\section{Introduction}
Deep reinforcement learning (RL) has accelerated at astounding speed in the last decade. Achieving superhuman performance in massively complex games such as Chess, Go \cite{silver2018general}, StarCraft II \cite{vinyals2019grandmaster}, Dota 2 \cite{berner2019dota}, and all 57 Atari games \cite{badia2020agent57}, deep RL has become a critical tool for Game AI. Many RL algorithms are benchmarked with games. The improvements in recent years in deep RL algorithms are often driven by the desire to improve the objective score performance of an agent and/or to reduce the training time or model size.  

Parallel to the impressive development of deep RL is the equally outstanding developments in quantum computing. Early theoretical work demonstrated the massive potential of quantum computers, such as Grover’s algorithm, which enables searching an unsorted list in $O(\sqrt{N})$ time \cite{grover1996fast}, and Shor’s algorithm which can break cryptosystems like RSA in polynomial time \cite{shor1999polynomial}. Only recently has quantum computation become more reality than spectre, with some claims of quantum supremacy, i.e. solving a problem that cannot be calculated on a traditional computer in any feasible amount of time \cite{arute2019quantum}. Quantum algorithms offer unique potentials because of their exploitation of quantum mechanical properties, such as superposition and entanglement (see the Quantum Computing section for more details).

Using quantum computing to help with machine learning tasks has attracted a lot attention in recent years. Quantum machine learning has significant potential to improve the speed of machine learning algorithms, with quantum perceptrons and quantum RL having theoretical potential for $O(\sqrt{N})$ speedups \cite{biamonte2017quantum}. Already work has been done to develop quantum GANs \cite{zoufal2019quantum} and quantum CNNs \cite{cong2019quantum}. Recently, the quantum RL field has been expanding with a variety of approaches such as using Grover Iterations \cite{ganger2019quantum} and CV photonic gates \cite{hu2019reinforcement} to solve gridworld environments. Other work has been done to envisage quantum computing as a RL problem \cite{khairy2020learning}. 

We explore the potential of utilizing quantum computing to aid with reinforcement learning tasks. We take inspiration from and extend the work done in \cite{yen2019variational} to use Quantum Variational Circuits (QVC), quantum circuits with gates parametrized by learnable values, in reinforcement learning. In \cite{yen2019variational}, QVCs were used with Double DQN to solve the deterministic 4x4 Frozen Lake OpenAI Gym environment. They reported that the parameter space complexity scales $O(N)$ in QVCs which is a significant improvement over the traditional neural network DQN which has parameter space complexity $O(N^2)$. However, their work only investigated QVCs with one algorithm that operated on a simple deterministic environment with a single input value and the observation space and output space we restricted to be the same number of quantum bits. We expand upon their work, evaluating more algorithms and multiple types of QVCs, creating new encoding schemes, and advancing to more complex environments. 

In this work, we use a quantum simulator to explore the potential for using quantum computing to solve reinforcement learning tasks. Expanding upon previous work, we present algorithms and encoding techniques that improve upon previous results. We apply our techniques to OpenAI Gym environments more complex than previous quantum RL work with largely positive results. Our results are indicative of the potential power of quantum computing in aiding reinforcement learning. 

\section{Reinforcement Learning}
The general formulation of reinforcement learning can be defined by an agent interacting with an environment attempting to maximize its reward function. This is often formulated as a Markov Decision Process (MDP). An MDP is characterized by the tuple $\langle S, A, P_a, R\rangle$, where S is the set of states, A the set of actions, $P_a$ is the probability of state transition $P_a = P[s_{t+1} = s^{\prime} | s_{t} = s, a_t = a]$, and R is the reward given for executing action $a_t$ at state $s_t$. In this work, our environments lack stochasticity and thus $P_a = 1$. The goal is to design an agent that operates policy $\pi$, $\pi(s_t) = a_t$, such that it maximizes the expected reward, $\mathbbm E [\sum_{t=0}^\infty R(s_t, a_t)|\pi]$. In learning the policy, the future reward is often discounted by a parameter $\gamma$.  

In order to learn $\pi$, deep RL relies on neural networks parametrized by weights and biases $\theta$. This paper relies on Q values estimation algorithms; the Q value being defined by $Q(s_t,a_t)=r_t+max_{a_{t+1}}Q(s_{t+1},a_{t+1})$, where $Q(s_t,a_t)$ is the Q value (or numerical estimation of reward) of taking action $a_t$ at state $s_t$, $r_t$ is the reward and $max_{a_{t+1}}Q(s_{t+1},a_{t+1})$ is the maximum future Q value. This Q function usually represented by a neural network. The general Q learning policy is defined for discrete actions spaces and can be formulated as such, $\pi(s;\theta) = max_aQ(s;\theta)$, i.e. the policy parametrized by $\theta$ is to choose the action that has the maximum Q value. To learn the policy utilizing neural networks a variation of the Bellman equation can be employed to calculate the mean squared loss function and from there, the gradients needed for backpropagation, $L_t(\theta)=\mathbbm E[(r_t+max_{a'}Q(s',a';\theta)-Q(s,a;\theta))^2]$ \cite{mnih2013playing}. This max operation can lead to over-estimations of the Q value, leading to convergence problems. A number of improvements to the vanilla DQN algorithm have been suggested. Double DQN \cite{van2016deep}, has a separate target network solely for predicting the future Q value inside the max operation, dueling DQN \cite{wang2016dueling} has separate network heads predict the advantage and value components of the Q value, distributional DQN \cite{bellemare2017distributional}, and noisy nets \cite{fortunato2018noisy}, to name a few. In this work the traditional DQN is used, to establish a base which variations can be applied and Double DQN because it does not require significant restructuring of the internals of the Q estimation function.

\section{Quantum Computing}
\subsection{Qubits and Superposition}
\begin{figure}[ht]
    \centering
    \includegraphics[scale=0.65]{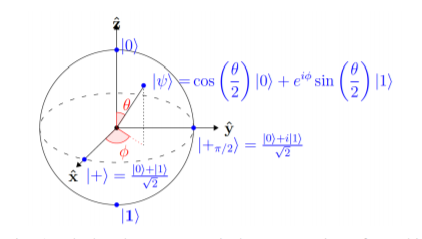}
    \caption{Bloch sphere representation of a qubit}
    \label{fig:1}
\end{figure}

The first important feature of quantum computing, critical to its representational and computational power, is the concept of superposition. In a classical computer, data is represented as a (binary) 0 or 1 and can be flipped between these states. The base unit of quantum computing is the quantum bit (qubit). Qubits rely on the quantum phenomenon of spin. The spin of a qubit is represented mathematically in the wavefunction. A quantum mechanical wavefunction, $\Psi$, represents the state of a system and can be a linear combination of components, e.g. $\Psi = \alpha|0\rangle + \beta|1\rangle$. These coefficients represent the probability amplitude of the wavefunction, i.e. $\int_{-\infty}^{\infty}|\Psi(x,t)|^2dx = 1$. Because a qubit can store information in this superposition, information representation scales with N qubits by $O(2^N)$ rather than $O(N)$ as with traditional computers. A single qubit can be visualized via a Bloch Sphere representing its wavefunction, see Figure \ref{fig:1} which is taken from \cite{cacciapuoti2020entanglement}. Classical binary states (i.e. 0 or 1) can be represented in the Z direction, $|0\rangle = \begin{bmatrix} 1 \\ 0 \end{bmatrix}$ represents spin up and $|1\rangle = \begin{bmatrix} 0 \\ 1 \end{bmatrix}$ represents spin down. However, as Figure \ref{fig:1} demonstrates, the state of the qubit can be anywhere on the sphere. Thus, this allows two states to be simultaneously represented in a 'superposition' (i .e. linear combination). Because the superposition is one of a probabilistic nature, when the measurement operator is applied the superposition collapses and only one state is measured, i.e. only a 0 or 1 is measured. 

\subsection{Gates}

In order to manipulate qubits, unique quantum gates must be used. There are many different quantum gates, but the ones most relevant to this work is the controlled NOT (CNOT) gate, Pauli gates and rotation gates. CNOT is important for its ability to induce entanglement. When qubits are entangled, they can no longer be represented as truly separate wavefunctions. Consider the two qubit entangled wavefunction $\Psi = \frac{|10\rangle+|01\rangle}{\sqrt{2}}$, i.e. an equal superposition of the states $|10\rangle$ and $01\rangle$. This wavefunction cannot be represented by two distinct single qubit wavefunctions, $\Psi= (p|0\rangle+q|1\rangle)(r|0\rangle+s|1\rangle)$, where $p,q,r,s$ are the coefficients that when squared yield the probability of measuring the qubit in that state. This would require $q*r = p*s = 1/\sqrt{2}$ and $p*r = q*s = 0$, which would require either p or r to be 0 making the first equation impossible. Thus, the qubits are called entangled because they are no longer isolated systems and share a wavefunction. When entangled actions done on either qubit will result in a change of the wavefunction, which will effect both qubits. Entanglement is important for quantum computing because it allows one operation to have an effect on multiple qubits in superposition, and doing the same on a classical computer would require many operations. CNOT is a two qubit gate and when acting on a purely spin up/spin down pair converts $|11\rangle$ into $|10\rangle$ and $|10\rangle$ into $|11\rangle$. However, when the qubits are not in pure spin up/down states, the wavefunction effects cannot be reflected in traditional computers (as changes in the wavefunction in superpositions are only allowed due to the quantum mechanical exploitation). The Pauli X, Y, Z gates flip the wavefunction about the specified axis. The rotation gates are a specialized version of the Pauli gates. he rotation gates are denoted: $R_x$, $R_y$, and $R_z$ and these gates rotate the qubit about the specified axis by the given $\theta$ radians.  

\subsection{Quantum Variational Circuits}
QVCs are a collection of gates that operate on a set of qubits \cite{benedetti2019parameterized}. They have a defined initial/input circuit, and a set of qubits and a collection of gates, parametrized by $\theta$, make up the body of the QVC. This collection of gates are denoted by $U(\theta)$. The parameters, $\theta$, is what is being 'learned'. For a given input, the circuit is evaluated and a 'readout operator' is applied to extract information from these gates as an output. The readout operator we use is the Pauli Z gate. I.e. the Pauli Z gate is applied which results in a numerical measurement. It is possible to apply other Pauli gates here and readout in that basis. We use the Pauli Z gate because Z, the 'computational basis' is common. External to the circuit a loss function and gradients are calculated in order to update the parameters. In this work, the loss function is the mean squared error. 

Calculating the gradient for a QVC requires different techniques than a neural network because the mathematical operations are fundamentally different. In this work the parameter shift differentiator is used. Although this is implemented by TensorFlow-Quantum the gradient calculations are an important difference between QVCs and deep neural networks, as such it is important to provide an adequate mathematical overview. In order to understand the gradient calculation formula we must understand both the individual gates and QVC as a function. The collection of gates $U(\theta)$ can be separated into a collection of $N$ layers operating on $M$ qubits. For a given layer $\ell$, it can be represented as a set of single qubit rotation gates operating in parallel, $U^{\ell}(\theta^{\ell}) = \bigotimes_{i=1}^M U_{i}^{\ell}(\theta_i^{\ell})$ \cite{broughton2020tensorflow}. Within this layer, each gate can enact a rotation of the qubit. This rotation of angles can be expressed similar to Euler's form: $U_{i}^{\ell}(\theta_i^{\ell}) = e^{-iaG\theta_i^{\ell}}$. Where $G$ is a linear combination of Pauli gates (called a generator). G can be represented as a $2 \times 2$ matrix (like all Pauli gates) and has two eigenvalues $e_0, e_1$ \cite{crooks2019gradients}. The derivative of this is straightforward, due to the nature of exponentials, $\pdv{}{\theta}U(\theta) = -iaGe^{-iaG\theta} = -iaGU(\theta)$. Prior to applying this differentiation we must present a big picture view of a QVC. A QVC is a function with parameters $\theta$. The output of this function is what the result of the Z gates. Thus, the QVC is a function that results in the expectation values from the Z gates. Prior to these Z gates being applied, however, the parametrized gates are applied. These parametrized gates change the starting wavefunction prior to the Z gates (as they are the very last gates used to generate output), or in quantum mechanical notation: $f(\theta) = \langle \Psi_0 | U^{\dag}(\theta) \hat{Z} U(\theta) | \Psi_0 \rangle$. The parameter shift rule states that $\pdv{}{\theta}f(\theta) = \langle \Psi_0 | (\pdv{}{\theta}U^{\dag}(\theta)) \hat{Z} U(\theta) | \Psi_0 \rangle + \langle \Psi_0 | U^{\dag}(\theta) \hat{Z} (\pdv{}{\theta}U(\theta)) | \Psi_0 \rangle$. We can combine this parameter shift rule with the derivative calculated above \cite{schuld2019evaluating}. This can then be reduced down to the final differentiating rule: $\pdv{}{\theta}f(\theta) = r[f(\theta+\frac{\pi}{4r}) - f(\theta-\frac{\pi}{4r})]$, where $r = \frac{a}{2}(e_1-e_0)$ \cite{crooks2019gradients}. This last equation is the parameter shift technique for how gradients are calculated for a QVC as seen in algorithm 1. 



\section{Approach}

In this work, we explore the potential for using QVCs in place of neural networks in RL algorithms. Substituting QVCs for neural networks requires almost no modification to the traditional algorithm. We evaluate four different variations, either being a pure QVC model or a hybrid QVC (i.e. a QVC which has outputs then fed into a single dense layer) in place of a neural network in either DQN and Double DQN. While all QVC are 'hybrid' in that they utilize traditional computers for loss calculations, we use the pure/hyrbid terminology to refer to whether or not the output of the QVC is fed into a single dense neural network layer. Hybrid QVCs have seen limited use in quantum RL but their positive results are important as they allow differences in the qubit observation and action spaces. We also modify the pure QVC technique to allow for differences in observation and action space by combining quantum pooling operations \cite{cong2019quantum} with traditional QVCs. We also experimentally evaluate QVCs representational power, as our results indicate that a QVC can perform comparably to neural networks with total parameters at least an order of magnitude larger. We present two new encoding schemes for different types of input data, both suitable for environments with arrays as inputs, an important advancement as many RL environments have more than single integer observation spaces. This work was done on simulations on a classical computer using the Noisy Intermediate Scale Quantum (NISQ) \cite{preskill2018quantum} simulator Cirq (from Google AI Quantum) and TensorFlow-Quantum (TFQ) \cite{broughton2020tensorflow}. 

\subsection{Quantum Data Encoding}
Special techniques are needed to work with classical data on a quantum device or simulator. Although one can convert all classical data to binary and represent that binary with the qubits, this is very inefficient, since a single precision floating point would then take 32 qubits (a substantial amount). While, theoretical encoding schemes do exist, e.g. an encoding scheme for arbitrary state preparation \cite{long2001efficient} and flexible representation of quantum images \cite{le2011flexible} these approaches are not yet suited for the use in QVC RL. The techniques in \cite{long2001efficient} are cost efficient, but require gates beyond TFQ's simulating power. And the technique from \cite{le2011flexible} uses available gates, but it required hundreds of gates which we found to be too high, making it impractical for data intensive applications like RL.

In order to solve the problems mentioned above we present two approaches to data encoding. They are both fast and effective, however, they are slightly below the theoretically optimal data representation. Although utilizing the same gates as \cite{yen2019variational}, these are fundamentally different algorithms. The technique presented in \cite{yen2019variational} converts a single integer into binary then uses that as input into the gates. Our algorithms can handle multiple inputs of both integers and floats (which are impractical to convert to binary). Our algorithms are also more efficient in terms of qubit usage, requiring $O(N)$ qubits (N = size of input array), superior to converting all numbers to binary as converting to binary would scale $O(Nlogn)$ with input (N = number of input elements, n = size of input integer). The first encoding scheme, which we call Scaled Encoding, is for environments that have input values with defined ranges. The process is simple, scale each input between 0 and $2\pi$ and rotate along $R_x$ and $R_z$ with the corresponding radians. This is fast (requiring only 2 gates per qubit) and is shown to be experimentally effective (see Blackjack section for more details). However, in some environments, e.g. CartPole, the range on some data points is between $-\infty$ to $\infty$ and the data is skewed such that even inserting artificial range cutoffs would be impractical. For environments such as this we use a different scheme, called Directional Encoding, defined by rotating each qubit $R_x$ and $R_z$ by either $\pi$ or 0 radians determined by the simple conditional: radians = $\pi$ if datapoint $>$ 0 else 0. I.e. if, and only if, the value is positive we rotate, specifically $\pi$ radians. This also only requires 2 gates per qubit and is shown to be experimentally viable. These encoding schemes scale $O(N)$ with the size of the input array, worse than the optimal $O(logN)$. These schemes represent an advancement over \cite{yen2019variational} because of their abilities of take more than 1 input value and improve on the representational complexity. 

\subsection{Model Architecture}
\begin{figure}[ht]
    \centering
    \includegraphics[scale=0.49]{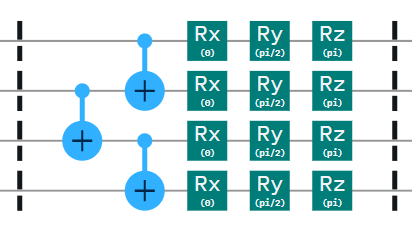}
    \caption{One 'layer' of the QVC, composed of CNOT and parametrized rotation gates}
    \label{fig:3}
\end{figure}

\begin{figure}[ht]
    \centering
    \includegraphics[scale=0.63]{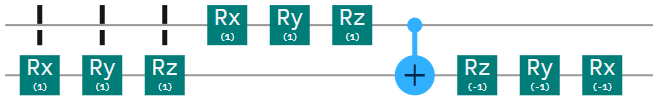}
    \caption{Quantum pooling operation}
    \label{fig:4}
\end{figure}

We implement and compare two versions of QVC models. In both cases, the body of the QVC is the same. It is composed of several ‘layers’ (they are called layers only because of visual aspects, the mathematical operations are not those of a neural network layer), seen in Figure \ref{fig:3} (Figure \ref{fig:3} and Figure \ref{fig:4} were generated with IBM Quantum Experience Circuit Composer). In this work, the QVCs use three layers. I.e. the block specified in Figure \ref{fig:3} is repeated 3 times. These are composed of $R_x$, $R_y$, and $R_z$ gates parametrized by different $\theta$. In addition, there is a collection of CNOT gates in front of the rotations whose primary goal is to entangle the qubits. 

In the hybrid model, the output is fed into a single dense output layer, one that has the same number of nodes as the action space. Thus, we reduce (or expand) the output to fit the action space. In pure QVC, we rely on quantum pooling techniques. The function of the quantum pooling operation is very similar to traditional pooling operations, it seeks to combine and reduce the size of layer so that the observation space can be reduced to the action space without losing information. Using the quantum pooling operation from \cite{cong2019quantum}, we can reduce the 2 qubits to 1 qubit, applying this operation as many times as needed. The pooling operation is defined parametrized $R_x$, $R_y$, $R_z$ gates, a CNOT gate, followed by the parametrized inverse rotation gates $R_x^{-1}$, $R_y^{-1}$, $R_z^{-1}$. Figure \ref{fig:4} shows a diagram of a single pooling operation where the (1) and (-1) represent the power. 

\subsection{QVC Versions of DQN and DDQN Algorithms}
The Double DQN and DQN algorithm are effectively the same as in \cite{van2016deep}. There are no algorithmic changes from the established DDQN algorithm. There are, however, necessary implementation differences. Prior to feeding the data into the QVC, the data must be encoded. The replay buffer functions in the same way as in traditional approaches, keeping track of the $\langle s, a, r, s^\prime \rangle$ tuples just with a different, encoded, state representation. We see this as one of the important facets of this work, the ease of implementation within existing algorithms. One does not have to fundamentally or drastically change an algorithm in order to apply the power of QVCs to it. The algorithm presented in Algorithm 1. 

\begin{algorithm}
Initialize replay buffer $\mathbbm D$

Initialize action QVC $\theta$, target QVC $\theta^t \xleftarrow{} \theta$

\For{episode = 0, N}{
  encode $s_1$ to quantum circuit $\xi_1 = \xi(s_1)$
  
  \While{game is not finished} {
  
    select action $a_t$ via $\epsilon$-greedy policy
    
    execute action $a_t$ and observe reward $r_t$ and next state $s_{t+1}$
    
    encode $s_{t+1}$, $\xi_{t+1} = \xi(s_{t+1})$
    
    store MDP tuple $\langle \xi_t, a_t, r_t, \xi_{t+1} \rangle$ in $\mathbbm D$
  }
  select random minibatch from $\mathbbm D$
  
  set $y_i = r_i + \gamma max_{a'}Q(\xi_{i+1},a';\theta^t)$
  
  calculate loss L according to $L(\theta) = (y_i - Q(\xi_i, a;\theta))^2$
  
  update parameters utilizing differentiator: $\pdv{}{\theta}f(\theta) = r[f(\theta+\frac{\pi}{4r}) - f(\theta-\frac{\pi}{4r})]$
  
  \eIf{episode mod C == 0}{
   $\theta^t \xleftarrow[]{} \theta$;
   }{
   $\theta^t = \tau \theta^t + (1-\tau)\theta$
  }
 }
 \caption{Q-DDQN}
\end{algorithm}

\section{Experiments}
All experiments were conducted with OpenAI Gym \cite{brockman2016openai}, specifically the CartPole and Blackjack environments. We chose these environments because they represent an advancement in complexity over previous research in quantum RL which has largely been dominated by gridworld environments, both in terms of policy complexity and input complexity (i.e. their input arrays are larger and have more possible values). They are also very different from each other, with different reward functions and strategies and utilize different encoding schemes. CartPole works with the Directional Encoding because the range of the input values is infinite. Blackjack utilizes the Scaled encoding scheme and demonstrates that this scheme is able to properly encode the magnitude of the data (which is lost for CartPole). This magnitude of the input is critical as the optimal strategy for playing blackjack with 1 point vs 20 points is substantially different (as if you have more than 21 points you bust and lose). While neither are as complex environments as StarCraft II or Dota 2, they demonstrate an notable advancement in complexity over previous work and are used as benchmarks for RL algorithms \cite{nagendra2017comparison}. Algorithm parameters were constant across experiments: initial $\epsilon$ of 1.0, $\epsilon$ decay of 0.9, $\epsilon$ minimum of 0.01, and a reward discount, $\gamma$, of 0.95. The optimizer, ADAM \cite{jlb2015adam}, and associated hyperparameters were constant for all experiments. The wall-clock training time of these on a single NVIDIA GTX 1070 GPU ranges from ~5-30 minutes.

\subsection{CartPole}
\begin{figure}[ht]
    \centering
    \includegraphics[scale=0.53]{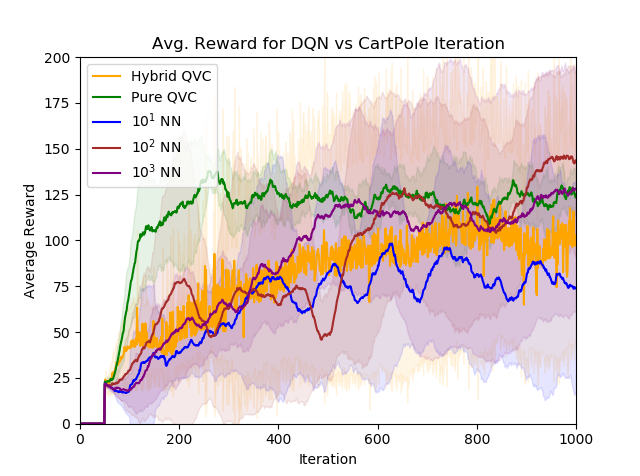}
    \caption{Comparison of NN and QVC DQN on CartPole}
    \label{fig:5}
\end{figure}

\begin{figure}[ht]
    \centering
    \includegraphics[scale=0.53]{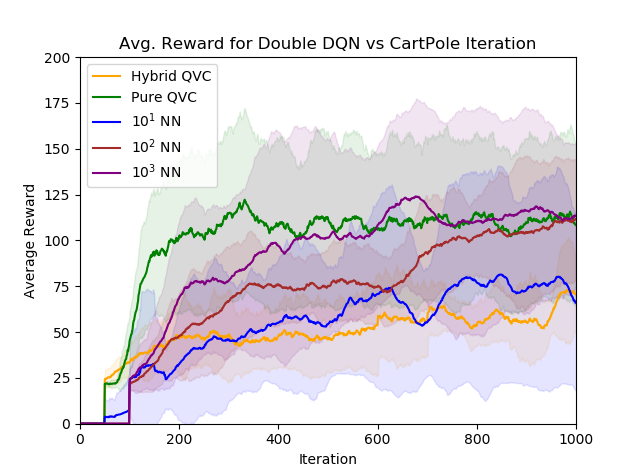}
    \caption{Comparison of NN and QVC DDQN on CartPole}
    \label{fig:6}
\end{figure}

The first environment is CartPole, which we use to compare QVC based DQN/DDQN with traditional deep neural network DQN/DDQN. The Directional encoding scheme is applied to both the neural network and the QVC. Specifically, this means that just as the qubits are encoded with 0s and 1s, so too does the neural network receive a binary array. All graphs begin at 50 iterations because the average reward is calculated by averaging over the previous 50 iterations. Figure \ref{fig:5} shows a comparison between the traditional neural network DQN and the two types (pure, hybrid) of QVC used. All shaded areas represent the 95\% confidence interval over 6 runs. This figure demonstrates that both hybrid and pure QVC models achieve a better policy and arrive at this policy faster than traditional neural networks. Figure \ref{fig:6} demonstrates the same comparison, using the Double DQN algorithm. This experiment demonstrates that the QVC models perform at least as well, if not better, than the neural network based models. 

Figure \ref{fig:5} and Figure \ref{fig:6} include neural networks with different numbers of parameters to show how well the QVCs perform in terms of representational abilities too. These figures show that QVCs have superior representational power over neural networks. In this work, the pure QVC has 48 trainable parameters in each 'network'. Each figure shows a comparison between the pure and hybrid QVC with parameter space of $10^1$ and neural networks with parameters on order $10^1, 10^2, 10^3$, specifically 58, 226, and 1,282 trainable weights (with 1, 2, 3 intermediate layers). This experimentally demonstrates the encoding and potential representational strength of QVCs, as they operate comparable to neural networks with orders of magnitude more parameters. 

\subsection{Blackjack}
\begin{figure}[ht]
    \centering
    \includegraphics[scale=0.53]{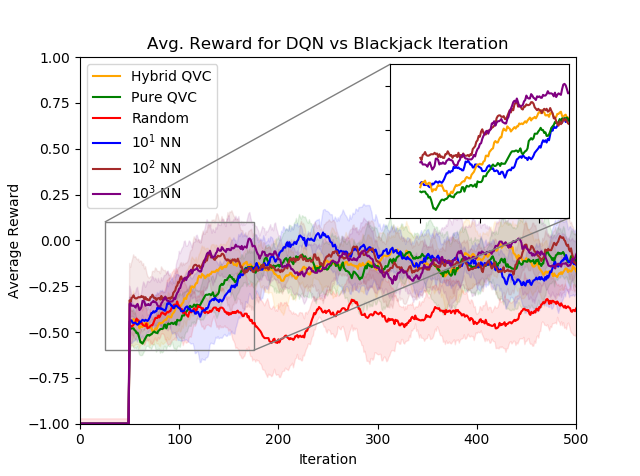}
    \caption{Comparison of NN and QVC DQN on Blackjack}
    \label{fig:7}
\end{figure}

\begin{figure}[ht]
    \centering
    \includegraphics[scale=0.53]{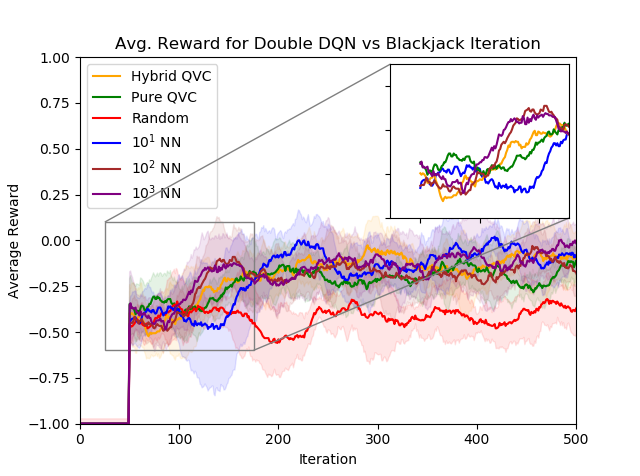} 
    \caption{Comparison of NN and QVC Double DQN on Blackjack}
    \label{fig:8}
\end{figure}

Blackjack's different reward approach stems from the fact that it originates as a casino game, designed such that the house always wins, i.e. the optimal policy is right below the 0 reward mark. This explains the discrepancy between the results in Figures \ref{fig:7} and \ref{fig:8} and for CartPole. Because the optimal policy is much more limited, and thus easier to attain, both pure and hybrid QVC achieve similar results. Figures \ref{fig:7} and \ref{fig:8} show comparisons of the DQN and DDQN algorithms, respectively. We compare the speed at which the model's learn is a metric by which comparisons can be made. The enlarged areas in Figures \ref{fig:7} and \ref{fig:8} demonstrate how the quantum approaches learn slightly faster than the same order parameter neural network approach (although slightly below the higher parameter networks). In this example the exact number of parameters are 33 for the QVC, and for the neural networks 38, 194, and 1,250 (with 1, 2, 3 intermediate layers). A random agent is also included to establish a baseline performance. This random agent is not included in CartPole as CartPole does not have as much inherent randomness that can cause random agents to perform at all comparably to RL agents and would be almost about 0.  This shows that the Scaled encoding scheme can be effective. 

\section{Discussion} 
\subsection{Differences in Hybrid and Pure}
While in three out of the four tests, the hybrid model performed comparably to the pure model, the Double DQN CartPole is the exception. This is slightly different in that the pure QVC model achieves a superior policy, whereas the hybrid performs similar to the neural network. This is the only case in which the discrepancy is significant. We suspect that hybrid models converge faster (for the early time steps the hybrid model is scoring better than the pure) and it is possibly converging on a local optimum rather than the global optimum policy because of its higher convergence speed. This faster convergence of the hybrid model becomes apparent in the Blackjack experiment. Because the optimal policy is less complex in this case, faster convergence becomes advantageous. 

\subsection{Generalizability}
This work suggests the potential for improved generalizability. Because the encoding schemes are more flexible and the observation and action spaces are no longer tied together, this should expand the problem space that these algorithms can be applied to. Naturally quantum data is also compatible with this approach. The encoding schemes should make the approach more general by allowing the input to be arrays of floating point values and integers. The Directional encoding scheme is designed to be applied to inputs that have an infinite range in which the magnitude does not matter, and the Scaled encoding scheme can be applied to inputs with well defined ranges that do not have a significant skew. This work is also generalizable to future improvements to the DQN algorithm, as any improvements made to the DQN algorithm can be utilized by this technique as well, e.g. prioritized experience replay \cite{DBLP:journals/corr/SchaulQAS15}. 

\subsection{Future Work}
Although this work suggests the potential for improved generalizability, further work is needed to verify this. To verify the generalizability, experiments should be conducted with more variations on the hyperparameters of the Quantum Variational Circuits and different applications. Expanding the applications of these algorithms to more complex environments (e.g. Atari) is a natural next step. In addition, we are interested in investigating more encoding schemes, as described in the Quantum Data Encoding Section. This is predicated upon the necessary gates becoming available in TensorFlow-Quantum. 

\section{Conclusion}
This work expands upon previous ideas and algorithms in the field of quantum computing and reinforcement learning to present Quantum Variational Circuit approaches to solve reinforcement learning tasks. We introduce and demonstrate the potential of both hybrid and pure Quantum Variational Circuits in both Double DQN and DQN algorithm variations using the CartPole and Blackjack OpenAI Gym environments. This work also demonstrate the potential of our two new classical to quantum data encoding schemes: Scaled encoding and Directional encoding. The success that these models achieved on both environments suggests that Quantum Variational Circuits may have representational abilities superior to traditional neural networks. This work is indicative of the potentially impactful relationship between quantum computing and reinforcement learning.


\end{document}